\documentclass{PoS}
\usepackage{amsmath,amssymb,amsfonts,amstext,amsthm}
\usepackage{algorithm,algorithmic}

\newcommand{\Dodwf}{\mathcal{D}}
\newcommand{\registered}{\textsuperscript{\textregistered \ }}
\newcommand{\TM}{\textsuperscript{TM\ }}

\title{GPU-Based Conjugate Gradient Solver for Lattice QCD with Domain-Wall Fermions}

\ShortTitle{CG for DWF on GPU}

\author{Ting-Wai~Chiu$^{1,2}$, 
        Tung-Han Hsieh$^{3}$, 
        \speaker{Yao-Yuan~Mao}$^{,1}$, 
        Kenji~Ogawa$^1$ (for the TWQCD Collaboration) \\
    $^1$ Department of Physics, and Center for Theoretical Sciences, National Taiwan University, Taipei 10617, Taiwan \\
    $^2$ Center for Quantum Science and Engineering, National Taiwan University, Taipei 10617, Taiwan \\
    $^3$ Research Center for Applied Sciences, Academia Sinica, Taipei 115, Taiwan
}

\abstract{We present the first GPU-based conjugate gradient (CG) solver for 
lattice QCD with domain-wall fermions (DWF). It is well-known 
that CG is the most time-consuming part in the Hybrid Monte Carlo 
simulation of unquenched lattice QCD, which becomes even more
computational demanding for lattice QCD with exact chiral symmetry. 
We have designed a CG solver for the general 5-dimensional DWF
operator on NVIDIA\registered CUDA\TM architecture with mixed-precision, 
using the defect correction as well as the reliable updates 
algorithms. We optimize our computation by even-odd preconditioning 
in the 4D space-time lattice, plus several innovative techniques 
for CUDA kernels.
For NVIDIA GeForce\registered GTX 285/480, our CG solver  
attains 180/233 Gflops (sustained).}

\FullConference{The XXVIII International Symposium on Lattice Field Theory, Lattice2010\\
		June 14-19, 2010\\
		Villasimius, Italy}

\begin{document}

\section{Introduction}

Simulation of unquenched lattice QCD with exact chiral symmetry is a grand challenge
among all sciences. Even for a modest $ 16^3 \times 32 $ lattice with lattice
spacing $ a \sim 0.1 $ fm, it often requires a supercomputer with peak computing power more than
50 Teraflops (e.g., 10 racks of IBM BlueGene/L). Therefore, only 2-3 lattice QCD
groups around the world could afford to perform the simulation of unquenched lattice QCD with
the domain-wall fermion \cite{Kaplan:1992bt}, or the overlap-Dirac fermion \cite{Neuberger:1997fp}.
However, this scenario has been undergoing a dramatic change during the last 12 months. 
With the emergence of low-cost and computationally powerful Graphic Processing Unit (GPU),
now plugging a graphic card with NVIDIA GTX 285 (240 cores, one Teraflops peak)
into a PC immediately turns the system into a powerful device, attaining a sustained
180 Gflops for the conjugate gradient (CG) with mixed precision.

Since 2009, the Taiwan Lattice QCD Collaboration (TWQCD) has been using a 
GPU cluster (currently constituting of 250 NVIDIA GPUs
with 40 Teraflops (sustained)) to simulate unquenched lattice QCD 
with optimal domain-wall quarks \cite{Chiu:2002ir, Chiu:2009wh}.
We have met the challenge of preserving the chiral symmetry 
to a high precision and sampling all topological sectors ergodically.
For our recent physical results on the 2-flavors QCD, we refer the 
readers to \cite{Chiu:2010ab} and \cite{Hsieh:2010ab}.

In this paper, we present our design of the first 
CG solver for the general 5-dimensional DWF operator 
on NVIDIA CUDA architecture with 
mixed-precision, using the defect correction 
as well as the reliable updates algorithms. 
Our CG solver is optimized with even-odd preconditioning 
on the 4-dimensional space-time lattice, plus
several innovative tuning techniques for the CUDA kernels.
For NVIDIA GeForce GTX 285/480, our CG solver  
achieves 180/233 Gflops (sustained).

\section{Conjugate Gradient Solver for Domain-Wall Fermions}

\subsection{Optimal Domain-Wall Fermions}

Mathematically, for a given $ N_s $ 
(the number of sites in the fifth dimension), 
the maximal chiral symmetry can be attained by 
the optimal domain-wall fermion (ODWF) \cite{Chiu:2002ir} with the operator
\begin{equation}
[\Dodwf(m_q)]_{xx';ss'} = 
  (\omega_s D_w + 1)_{xx'} \delta_{ss'}
 +(\sigma_s D_w - 1)_{xx'} L_{ss'}, \quad (\sigma_s = \omega_s) 
  \label{eq:D_odwf_def}
\end{equation}
Here $D_w$ is the standard Wilson Dirac
Operator plus a negative parameter $-m_0 \; (0 < m_0 < 2)$,
\begin{equation}
(D_w)_{xx'} = -\frac{1}{2} \sum_{\mu} \left[
  (1-\gamma_\mu)U_\mu(x)\delta_{x+\hat{\mu},x'}
 +(1+\gamma_\mu)U^\dagger_\mu(x')\delta_{x-\hat{\mu},x'} \right]
 + (d - m_0),
\end{equation}
where $U_\mu(x)$ denotes the link varaible, $d$ is the dimension 
of the space-time ($d=4$ for QCD),   
\begin{equation}
L = P_+ L_+ + P_- L_-, \quad P_\pm = (1\pm \gamma_5)/2,
\end{equation}
and
\begin{equation}
(L_+)_{ss'} = \left\{ 
    \begin{array}{ll} \delta_{s-1,s'}, & 1 < s \leq N_s \\ 
        -(m_q/2m_0) \delta_{N_s,s'}, & s = 1 \end{array}\right.,
\quad L_-=(L_+)^{T}.
\end{equation}
The weights $ \{ \omega_s \} $ along the fifth dimension are 
fixed according to the formula derived in \cite{Chiu:2002ir} such 
that the maximal chiral symmetry is attained. 
In general, for other DWF with non-maximal chiral symmetry, 
the weights $\{\sigma_s\}$ and $\{\omega_s\}$ have 
different values, e.g., for the conventional (Shamir) DWF, 
$\sigma_s = 0, \omega_s = 1, \forall s $,
and for the Borici DWF \cite{Borici:1999zw}, 
$\sigma_s = \omega_s = 1, \forall s $.

\subsection{Even-Odd Preconditioning}

Since $D_w$ commutes with $(\omega)_{ss'} \equiv \omega_s \delta_{ss'}$ and 
$(\sigma)_{ss'} = \sigma_s \delta_{ss'} $, Eq.~(\ref{eq:D_odwf_def}) becomes 
\begin{equation}
\Dodwf(m_q)=D_w(\omega+\sigma L)+(1-L).
\label{eq:D_odwf_def2}
\end{equation}
Separating the even and the odd sites on the 4D space-time lattice, 
Eq.~(\ref{eq:D_odwf_def2}) can be written as
\begin{equation}
\Dodwf(m_q)=
\begin{pmatrix}
d - m_0 & D_w^{\text{EO}} \\
D_w^{\text{OE}} & d - m_0 
\end{pmatrix} 
(\omega+\sigma L)+(1-L) 
=
\begin{pmatrix}
X & D_w^{\text{EO}} Y \\
D_w^{\text{OE}} Y & X
\end{pmatrix},
\label{eq:D_odwf_eo}
\end{equation}
where
\begin{equation}
X \equiv (d - m_0 )\omega(1+c L)+(1-L),\quad Y \equiv \omega(1 + c L), 
\quad  (c)_{ss'} \equiv (\sigma_s/\omega_s) \delta_{ss'}.
\end{equation}
Now we further rewrite it in a more symmetric form by defining
\begin{equation}
  \label{eq:m5}
M_5\equiv \sqrt{\omega}^{-1}YX^{-1}\sqrt{\omega}
 = \left[(d-m_0)
 + \sqrt{\omega}^{-1}(1-L)(1+cL)^{-1}\sqrt{\omega}^{-1}\right]^{-1},
\end{equation}
and
\begin{equation}
S_1\equiv \sqrt{\omega}^{-1}YX^{-1} = M_5 \sqrt{\omega}^{-1},
\quad S_2\equiv Y^{-1}\sqrt{w}.
\end{equation}
then Eq.~(\ref{eq:D_odwf_eo}) becomes
\begin{equation}
\Dodwf(m_q)=
S_1^{-1}
\begin{pmatrix}
1 & M_5 D_w^{\text{EO}}  \\
M_5 D_w^{\text{OE}} & 1 
\end{pmatrix}
S_2^{-1}
=S_1^{-1}
\begin{pmatrix}
1 & 0 \\
M_5 D_w^{\text{OE}} & 1
\end{pmatrix}
\begin{pmatrix}
1 & 0 \\
0 & C
\end{pmatrix}
\begin{pmatrix}
1 & M_5 D_w^{\text{EO}} \\
0 & 1
\end{pmatrix}
S_2^{-1},
  \label{eq:D_odwf_decomp}
\end{equation}
\begin{equation}
\label{eq:c_def}
C \equiv 1 - M_5 D_w^{\text{OE}} M_5 D_w^{\text{EO}}.
\end{equation}
Obviously, the most time-consuming task in the HMC  
is to solve the linear system  
$ C C^\dagger |x\rangle = |b\rangle $ by 
the conjugate gradient (CG), namely,   
in the computation of the fermion force in the molecular dynamics. 
In this work, we implement the entire conjugate gradient
inside the NVIDIA GPU, which can be used for the HMC as well as      
for computing the valence quark propagators.

\subsection{Algorithm}

Conjugate Gradient (CG) method \cite{Hestenes} is a widely-used 
numerical algorithm for iteratively solving a 
linear system $Ax=b$ to a certain precision $\varepsilon$, where 
$A$ is a positive-definite Hermitian matrix. 
With the CG algorithm (see Algorithm 
\ref{alg:cg}), the problem is turned into a task dominated by
the matrix-vector multiplication. In this work, we utilize CUDA 
to implement the 5D domain-wall fermion operator  (\ref{eq:D_odwf_decomp})
matrix-vector multiplications of the CG on NVIDIA GPUs.
\begin{algorithm}
\caption{Conjugate Gradient}
\label{alg:cg}
\begin{algorithmic}
\STATE $ x_0 := $ initial guess
\STATE $ r_0 := b- A x_0 $
\STATE $ p_0 := r_0 $
\STATE $ k := 0 $
\WHILE{$\lvert r_k \rvert > \varepsilon \lvert b \rvert $}
\STATE $ \alpha_k := \left(r_k, r_k \right)/\left(p_k, A p_k \right) $
\STATE $ r_{k+1} := r_k - \alpha_k A p_k $
\STATE $ \beta_{k+1} := \left(r_{k+1}, r_{k+1} \right)/\left(r_{k}, r_{k} \right) $
\STATE $ x_{k+1} := x_k + \alpha_k p_k $
\STATE $ p_{k+1} := r_{k+1} + \beta_{k+1} p_k $
\STATE $ k := k + 1 $
\ENDWHILE
\end{algorithmic}
\end{algorithm}
For the GPU, the single-precision operations are several times 
faster than the double-precision ones, thus it is advantageous
to use the mixed-precision CG. In the so-called
\textit{defect correction} algorithm, 
one solves $ x $ in the low-precision,  
and updates the solution $ \hat{x} $ and the residue $ \hat{r} $ in the high-precision. 
(see Algorithm \ref{alg:cg_mp}, where the hatted symbols represent 
variables in the high-precision). 
In this fashion, most of the floating-point operations are in 
the low-precision, thus it is advantageous for the GPU.
\begin{algorithm}
\caption{Mixed-Precision Conjugate Gradient (Defect Correction)}
\label{alg:cg_mp}
\begin{algorithmic}
\STATE $ \hat{x} := $ initial guess
\STATE $ \hat{r} := \hat{b} - \hat{A} \hat{x} $
\WHILE{$\lvert \hat{r}_k \rvert > \hat{\varepsilon} \lvert \hat{b} \rvert $}
\STATE $ r := \hat{r} $
\STATE $ p := \hat{r} $
\STATE $ x := 0 $
\STATE Use Algorithm \ref{alg:cg} to solve $ x = A^{-1} r $ in the low-precision 
       to a precision $\varepsilon$
\STATE $ \hat{x} := \hat{x} + x $ 
\STATE $ \hat{r} := \hat{b} - \hat{A} \hat{x} $
\ENDWHILE
\end{algorithmic}
\end{algorithm}
Theoretically, the \textit{defect correction} algorithm is mathematically sound, 
and it always works in practice.  
However, the seemingly drawback is that one has to build up the Krylov space 
every time it restarts the CG in the low precision. 
On the other hand, if one does not reset the low-precision 
$p$ vector inside the while loop of Algorithm \ref{alg:cg_mp} 
(i.e., skipping the step ($ p:= \hat{r} $) except at the first time),
the ``warm-up" time in re-building the Krylov space 
could be reduced.
This so-called \textit{reliable updates} algorithm 
\cite{Sleijpen:96,Clark:2009wm}
would run faster than the \textit{defect correction}. 
Although the reliable updates in this fashion 
may not converge for all cases  
due to the non-orthogonality of $p$ and $r$,
in practice it seems to work for most cases.
We have implemented both 
algorithms in our CG solver, 
and it automatically switches to the \textit{defect correction} 
if the \textit{reliable updates} does not converge in the first place.

\section{CUDA Kernels and Optimization}

The CUDA architecture developed by NVIDIA enables us
to do parallel computations on NVIDIA's GPUs. 
(For more detailed programming models and strategies,
see ``CUDA Programming Guide for CUDA Toolkit'' \cite{cuda}.)

In CUDA, a \textit{thread} is the smallest unit
to execute a \textit{kernel}, and a certain number of threads form 
a \textit{block}. Each thread will be assigned a 
3-dimensional index, and each block a 2-dimensional index. 
Inside a block, a \textit{warp} of threads will execute the
\textit{kernel} concurrently. Each thread has its own \textit{register} memory, 
and threads in the same block share a \textit{shared} memory. The space of
register and shared memory is very limited, but they have the 
fastest access speed. The \textit{global} memory on the device
can be accessed by any thread, but it has the slowest bandwidth.

To implement the mixed-precision CG for ODWF with CUDA,
we perform all matrix-vector multiplication, vector reduction 
(inner product), and vector addition/subtraction on the GPU (device), 
while the CPU (host) is used to do the flow control, memory 
assignment, and device control. 

The CUDA kernels in our CG solver can be divided 
into five different catalogs. We will discuss each catalog 
and their optimization in the following subsections.

\subsection{Vector Index Conversion}

These kernels are used to change the indexing 
schemes between the device (GPU) and the host (CPU). 
To store the Dirac spinors in an
one-dimensional array, we need to map the 
multi-dimensional indices to the 1D array index. 
One needs $4 \times 3 \times 2 = 24$ real numbers to store
one Dirac spinor at each site of the 5D lattice.
On the CPU, this color-spinor index $c$ which
runs from $0$ to $23$ is the inner-most 
(fastest-running) index, which is followed by the fifth-dimension index 
$s$, and then $ x, y, z, t $ indices, where $t$ is the outer-most 
(slowest-running) index. If $i_{host}$ denotes the one-dimensional 
array index of the CPU scheme, then we have
\begin{equation}
i_{host} = c + s \times 24 + x \times 24 N_s  + \cdots 
           + t \times 24 N_x N_y N_z N_s.
\end{equation}
However, for computation on the GPU, we assign each
thread a five-dimensional site index. This implies that 
adjacent threads have consecutive $s$ indices. Thus we want to 
arrange the data such that optimal coalescence is attained 
when loading the vector from device global memory to 
the register and/or the shared memory of the threads. 
Since the GPU provides vector data types such as {\tt float4} and 
{\tt double2} which allow us to move 4 single precision numbers (or 2 double precision
numbers) at one time, a simple way to map to the one-dimensional array 
index on GPU (for single-precision) is 
\begin{equation}
i_{dev} = c\hspace{-0.6em}\mod 4 + s \times 4 + [c/4]\times 4N_s 
          + x \times 24 N_s  + \cdots + t \times 24N_xN_yN_zN_s,
\end{equation}
and similarly for the double-precision.
So every time when we transfer data between the host and the device, 
we convert the index accordingly.

\subsection{Matrix-Vector Multiplication for $D_w^{\text{OE}} (D_w^{\text{EO}})$ }

$D_w^{\text{OE}} (D_w^{\text{EO}})$ 
is similar to the usual Wilson-Dirac operator $D_w$ without the mass term,
\begin{equation}
  \label{eq:Dwoe}
\left[ D_w^{\text{OE}} (D_w^{\text{EO}}) \right]_{xx'} 
= -\frac{1}{2} \sum_{\mu} \left[
  (1-\gamma_\mu)U_\mu(x)\delta_{x+a\hat{\mu},x'}
 +(1+\gamma_\mu)U^\dagger_\mu(x')\delta_{x-a\hat{\mu},x'} \right].
\end{equation}
From this expression we see that the multiplication of 
$D_w^{\text{OE}} (D_w^{\text{EO}})$ 
with a vector involves the link variables 
$U_\mu(x)$ and $\gamma$-matrices. We have used the following tricks 
for optimization.

Firstly, since the $\gamma$-matrices in Eq.~(\ref{eq:Dwoe}) 
are in the combination $(1 \pm \gamma_\mu)$, the left-handed
and the right-handed Dirac components are related to each other. 
Also, since the link variables do not have 
Dirac indices, we can just multiply $U_\mu(x)$ to 
the left-handed components,  
and then restore the right-handed components.

Secondly, $U_\mu(x)$ has no fifth-dimension dependence, so 
threads having the same $ x $ but different
$s$ can share the same $U_\mu(x)$. 
So we put the link variables in the shared memory.

Thirdly, because GPU computation is memory bandwidth bound, 
one must try to reuse the data. For example, the hopping term 
($\delta_{x-a\hat{\mu},x'}$) in $D_w$,    
all neighboring sites of $ x $ are involved in the calculation.
If we assign each $ x $ to one thread, then  
there must be overlapping data loading for neighboring sites.
To reduce this overlapping data transfer,  
we distribute each $(x, y, z)$ to one thread, 
with a loop in the $t$-direction. Then the 
neighboring data in the $t$-direction can be reused,  
and the efficiency is enhanced.

Besides above tuning techniques, we also expand small loops, 
and to use the texture memory for caching data, 
Here texture is used for loading the vectors and 
link variables. We use Python \cite{python} 
to expand small loops, 
and to generate the set of kernels for 
$ D_w $ multiplication.

\subsection{Matrix-Vector Multiplication for $M_5$}

The matrix $M_5$ is given in Eq.~(\ref{eq:m5}). One can see that
$M_5$ is block diagonal in the chiral basis and it
does not depend on the space-time nor the color indices.
In fact, it can be divided into two constant matrices in the fifth
dimension, i.e., the left-handed and the right-handed ones. 
So the multiplication of $M_5$ with a vector can be regarded as  
$ u_s = \sum_{s'} (M_5)_{ss'} v_{s'}$.   
Here we use the shared memory to store the source vector ($v$). 
Since $M_5$ only takes $2N_s^2$ real numbers, 
we can put $M_5$ into the register of each 
thread (with different $s$ and $x, y, z$).
Again, a loop in $ t $ is inserted to   
reuse the $M_5$ data in the register.
Also, we use Python to generate these kernels.

\subsection{Vector Reduction}

To calculate the norm of a vector, we use the well-known 
\textit{parallel reduction} with the shared memory.
However, due to the limitation on the number of threads per block, 
it is inevitable to use global memory when the size of a vector
becomes very large. Our solution is to perform the  
\textit{block reduction} in prior kernels, i.e., 
to sum up vector elements
(already stored in registers/shared memory) within each block.
Then these partial sums can be added with a parallel reduction.

\subsection{Vector Addition and Subtraction}
We can combine the simple
addition/subtraction with other kernels in which 
one vector has been loaded. 
For example, to multiply  
$C \equiv 1 - M_5 D_w^{\text{OE}} M_5 D_w^{\text{EO}} $ to a vector, 
we can combine the last subtraction with the last $M_5$ multiplication.

\section{Performance}

We present some benchmarks of our CG solver, using 
NVIDIA GeForce GTX 285, GeForce GTX 480, Tesla\TM C1060,
and Tesla C2050. 
Note that our code has not yet been well-tuned for the Fermi 
architecture (GTX 480 and C2050). 
From Table \ref{table:device_comp}, we see that the bottleneck of our 
program is in the single-precision $D_w$ matrix-vector multiplication. 
Due to the mixed-precision CG, the time used in the 
double-precision operations are almost negligible. 
For the Fermi architecture, due to the larger L1 cache, 
there is a significant improvement in the single-precision $D_w$ matrix-vector 
multiplication, and also in the double-precision $M_5$ matrix-vector 
multiplication. 
\begin{table}[htbp]
  \caption{Benchmark of our CG solver for DWF on a $16^3 \times 32 \times 16$ lattice, 
           numbers in units of Gflops)}
  \label{table:device_comp}
  \begin{center}
  \begin{tabular}{| l || r | r | r | r | r |}
\hline 
  & $D_w$ (single) & $M_5$ (single) & $D_w$ (double) & $M_5$ (double) & CG (mixed) \\ \hline \hline
GTX 285 & 177 & 346 & 33 & 69 & 181 \\ \hline
GTX 480 & 248 & 331 & 32 & 116 & 233 \\ \hline
C1060 & 128 & 290 & 29 & 61 & 132 \\ \hline
C2050 & 160 & 239 & 22 & 100 & 156 \\ \hline
  \end{tabular}
  \end{center}
\end{table}

\section{Summary}
We have implemented an efficient GPU-based CG solver for 
generalized domain-wall fermions in lattice QCD. 
Our CUDA kernels are tuned with several innovative techniques. 
On NVIDIA GeForce GTX 285/480, 
our CG solver achieves 180/233 Gflops (sustained). 
This efficient CG solver constitutes the most crucial part 
in TWQCD's HMC code for simulation of unquenched lattice QCD 
with the optimal domain-wall fermion.

\begin{acknowledgments}
  This work is supported in part by  
  the National Science Council 
  (Nos. NSC96-2112-M-002-020-MY3, 
        NSC99-2112-M-002-012-MY3, NSC96-2112-M-001-017-MY3, NSC99-2112-M-001-014-MY3, NSC99-2119-M-002-001) 
  and NTU-CQSE~(Nos. 99R80869, 99R80873).
\end{acknowledgments}

\end{document}